\documentclass[article]{aa} 
\usepackage{graphicx}

\def\x2{$\chi^{2}$}

\def\ginga{{\it Ginga}}

\def\xmm{{\it XMM-Newton}}
\def\chandra{{\it Chandra}}

\def\lunits{$\rm erg~s^{-1}$}
\def\funits{$\rm erg~cm^{-2}~s^{-1}$}
\def\cunits{$\rm cm^{-2}$}

\begin{document}

  \title{Can photo-ionisation explain the decreasing fraction of X-ray obscured AGN with luminosity?}

   \titlerunning{The fraction of X-ray obscured AGN}
    \authorrunning{A. Akylas \& I. Georgantopoulos}

   \author{A. Akylas and  I. Georgantopoulos}

   \offprints{A. Akylas}

   \institute{Institute of Astronomy \& Astrophysics, National Observatory 
 of Athens, I.Metaxa \& B. Pavlou, Penteli, 15236, Athens, Greece \\
     }   
       \date{}

   \abstract
   {{\chandra} and {\xmm} surveys show that the fraction  of obscured 
   AGN decreases rapidly with increasing luminosity. Although this 
   is usually explained by assuming that  the covering factor of 
   the central engine is much smaller at luminous QSOs, the exact 
   origin of this effect remains unknown.  We perform toy 
   simulations to test whether photo-ionisation of the obscuring screen 
   in the presence of a strong radiation field  can reproduce this effect. 
   In particular, we create X-ray spectral simulations using  a warm 
   absorber model assuming a range of input column densities and 
   ionisation parameters. We fit instead the simulated spectra with a 
   simple cold 
   absorption power-law model that is the standard practice in X-ray surveys.  
   We find that the fraction of absorbed AGN should fall with luminosity as 
   $L^{-0.16\pm0.03}$ in rough agreement with the observations. 
   Furthermore, this apparent decrease in the obscuring material is 
   consistent with the dependence of the FeK$\alpha$ narrow-line equivalent 
   width on luminosity, ie. the X-ray Baldwin effect. 
   {}}

   \maketitle

\section{INTRODUCTION}
 
The deepest {\chandra} X-ray surveys find a high surface density 
of  AGN (Alexander et al. 2003, Giacconi et al. 2002). At the 
faintest fluxes probed, $\rm f_{X}(2-10 keV)\approx 10^{-16}$ \funits,
the large majority of these sources ($\sim$ 80\%) are obscured, 
having a hydrogen column density of $\rm N_H>10^{22}$ \cunits 
(e.g. Akylas et al. 2006; Tozzi et al. 2006). 
The high fraction of obscured AGN witnessed by {\chandra} is well in 
agreement with previous X-ray background synthesis models 
(e.g. Comastri et al. 1995, 2001).  
However, the situation is drastically different at the brighter fluxes 
probed by the {\xmm} surveys. These show a marked deficit 
of X-ray obscured AGN (Piconcelli et al. 2002;  Perola et al. 2004;
Georgantopoulos et al. 2004;  Caccianiga et al. 2004).
The fraction of the obscured AGN drops to $\sim30\%$ at fluxes 
of $\rm f_{X}(2-10~keV)\approx 10^{-14}$ \funits.
In flux limited samples, a small increase in the obscured AGN fraction 
with decreasing observed flux is expected, as the absorbed sources 
 present lower flux. However the observed increase is 
much steeper than that predicted (eg. Piconcelli et al. 2002).

This discrepancy has been resolved with recent {\chandra} and {\xmm}  
observations (e.g. Akylas et al. 2006; La Franca et al. 2005; Ueda et al. 2003) 
which demonstrate that the X-ray obscured AGN fraction is lower at higher 
luminosities. 
This  can also explain the deficit of type-2 narrow-line 
QSOs at high redshift (Steffen et al. 2003). Recent X-ray background 
synthesis models that take into account the dependency of the obscured 
AGN fraction on luminosity (Gilli, Comastri \& Hasinger 2007) are very 
successful in predicting the observed flux distribution.

The important question is what is the physical mechanism 
that can produce the decrease in the fraction of obscured AGN. 
It is possible that in the presence of a strong radiation 
field the fraction covered by the torus may be reduced. 
K\"{o}ningl \& Kartje (1994) for example proposed a disc-driven 
hydromagnetic wind model of the torus where for high bolometric 
luminosities the radiation pressure is expected to flatten the torus. 
A similar model, the receding torus model,  has been invoked by  
Lawrence (1991) and Simpson  (2005), to explain the fraction of 
radio AGN optically classified as type-2. 
In the presence of a strong radiation 
field the dust is heated and sublimates when it reaches a temperature of 
about 1500-2000K. At high luminosities the sublimation radius increases and 
thus effectively the torus opening angle increases, resulting 
in a higher number of type-1 AGN.  

A similar effect is expected to take place in X-ray wavelengths. 
The X-ray absorbing gas is situated close to the nucleus as 
inferred from X-ray variability studies of type-2 AGN (Risaliti, 
Elvis \& Nicastro 2002). Thus it is reasonable to assume that a 
strong radiation field will also ionise an appreciable fraction of the 
surrounding obscuring screen, resulting in the decrease in the 
``effective'' column density. In this paper, we perform toy X-ray spectral 
simulations to estimate the strength of this effect. 
We also discuss any possible connection with the X-ray Baldwin effect.

\section{X-ray Simulations}

\subsection{Simulation Description}  

We assume a constant $n_er^2$ relation for all AGN, where $n_e$ and $r$ 
are the  density of the obscuring screen and its distance 
from the nucleus respectively. 
This choice ensures a uniform ionisation state for the circumnuclear 
material i.e. the ionisation parameter $\xi$(=L/n$_e$r$^2$ ergs cm s$^{-1}$) 
is the same at all distances from the nucleus. 
This could also imply that at a given distance from the central Black Hole the 
density is roughly the same in all AGN. 
Although this may not be the case, it is indirectly supported by the observational 
estimates for the density of the Broad Line Region (BLR) suggesting a uniform value 
of $\sim$10$^9$ cm$^{-3}$ (Peterson 2006) across a large number of nearby AGN. 

We use the {\sl XSPEC} v12.0 package (Arnaud 1996) 
to generate 1000 simulated {\chandra} spectral files according to the 
following specifications: 
\begin{enumerate}
\item{A power-law model modified by an ionised absorber ({\sl ABSORI} 
 model, Done et al. 1992) with a photon index fixed to 1.9 
(Nandra \& Pounds 1994).} 
\item{A redshift of 0.8 that corresponds to the peak redshift 
 in the deep Chandra X-ray fields (e.g. Tozzi et al. 2006)
 and thus the average redshift of the sources that produce the 
 X-ray background.} 
\item{An  amount of $N_H$ randomly varying in the range  
of $10^{21}-10^{24}$ cm$^{-2}$. We use  a  ratio of obscured 
($N_H>10^{22}$ cm$^{-2}$ ) to unobscured ($N_H<10^{22}$ cm$^{-2}$ ) 
sources of 4:1 that is similar to that  observed in the {\chandra} deep fields 
 at the lowest flux levels (e.g. Akylas et al. 2006, Tozzi et al. 2006).
 We further assume, a flat distribution of $N_H$ in the 
 10$^{22}$ to 10$^{24}$ cm$^{-2}$ region.}  
\item{An ionisation parameter,  $\xi$=L/n$_e$r$^2$ ergs cm s$^{-1}$, 
 (where L is the source luminosity in the 13.6 eV to 13.6 keV bandpass),
 randomly varying between 0 and 1000.} 
\item {An arbitrary combination of an intrinsic 2-8 keV flux of $\sim10^{-14}$ {\funits} and  an 
 exposure time of 50 ks to ensure a minimum of 10 counts per spectrum regardless of the $N_H$ and $\xi$ values.}
\end{enumerate}

\begin{figure}
   \centering
   \includegraphics[width=8.0cm]{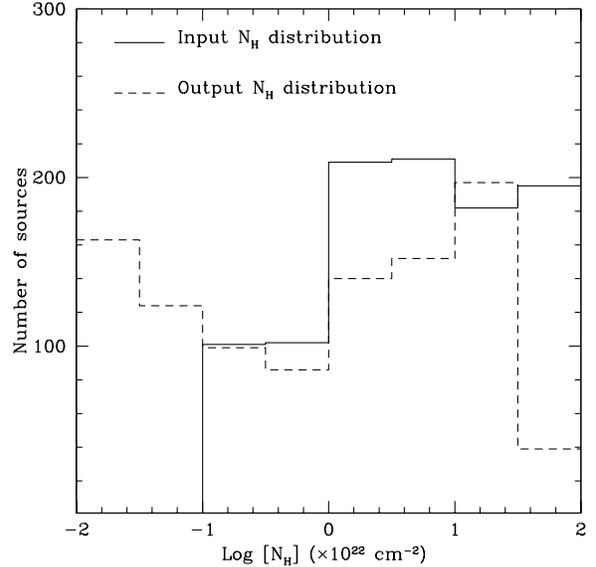}
     \caption{The solid line histogram describes the input $N_H$ 
distribution used for the warm absorption simulations and the 
short dashed line the output one, after fitting the simulated 
data with a cold absorption power-law model.}
    \label{nh_sim}
    \end{figure}

These files are then fitted in {\sl XSPEC}, using a power-law model 
modified by a cold photoelectric absorption model ({\sl WA*PO})
to estimate the ``effective'' neutral column density. 
We consider only the 0.3-8 keV energy band in the spectral fits 
where the {\chandra} effective area is high. Sources with adequate photon 
statistics ($>$ 100 counts)  are fitted using  the $\chi^2$ statistic  while for sources with less than 
100 counts we use the most appropriate C-statistic technique (Cash 1979).
We compare the parent and measured column density distribution in 
Fig. \ref{nh_sim}. 
The solid line histogram describes the input $N_H$ distribution 
assumed in the simulation ({\sl ABSORI*PO} model) and the short dashed 
line the output one ({\sl WA*PO} model). 
Note that due to the poissonian errors in each $N_H$ bin the shape of 
input $N_H$ slightly deviates from the assumed one.
Clearly, there is a significant 
reduction in the number of obscured ($N_H>10^{22}$  cm$^{-2}$) 
sources in comparison to the initial distribution.

\subsection{The simulated spectra}  
An important question that arises is whether the typical X-ray surveys
such as carried out by {\chandra} or {\xmm} can discriminate between a 
photo-ionised (i.e. {\sl ABSORI} model) 
and a neutral absorption model (i.e. {\sl WA} model)  
given the observational limitations i.e. the limited signal-to-noise ratio 
and energy resolution.
The answer to this question depends primarily on the spectrum quality
in the sense that in the sources with many counts we can more easily 
reject the neutral absorption model based on the $\chi^2$ statistic result.
The amount of the $N_H$ and its degree of ionisation $\xi$ also play an important role 
given that their value may cause  more or less prominent ionisation features 
to appear in the X-ray spectrum. 

\begin{figure}
   \centering
   \rotatebox{0}{\includegraphics[width=8cm, height=8cm]{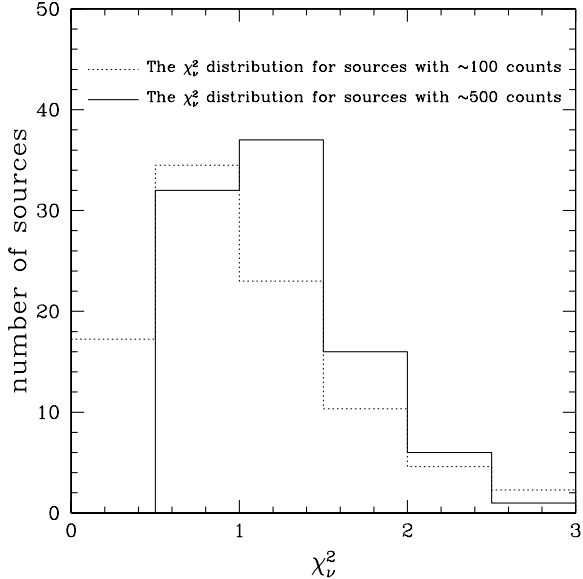}}
     \caption{The distribution of the  $\chi^{2}_{\nu}$ values for the ``low count'' 
       (dotted line) and the ``high count'' (solid line) samples.}
    \label{chi}
    \end{figure}

To explore this we examine two different samples, each one representative of our simulation, 
containing $\sim$100 sources each. These sets present  $\sim$100 and $\sim$500 counts  
respectively in their 0.3-8 keV spectrum. In Fig. \ref{chi} we present the distribution 
of the best reduced $\chi^2$ values ($\chi^{2}_{\nu}$) when fitting the data using a cold absorption power-law 
model with $\Gamma$ fixed to 1.9.

\begin{figure*}
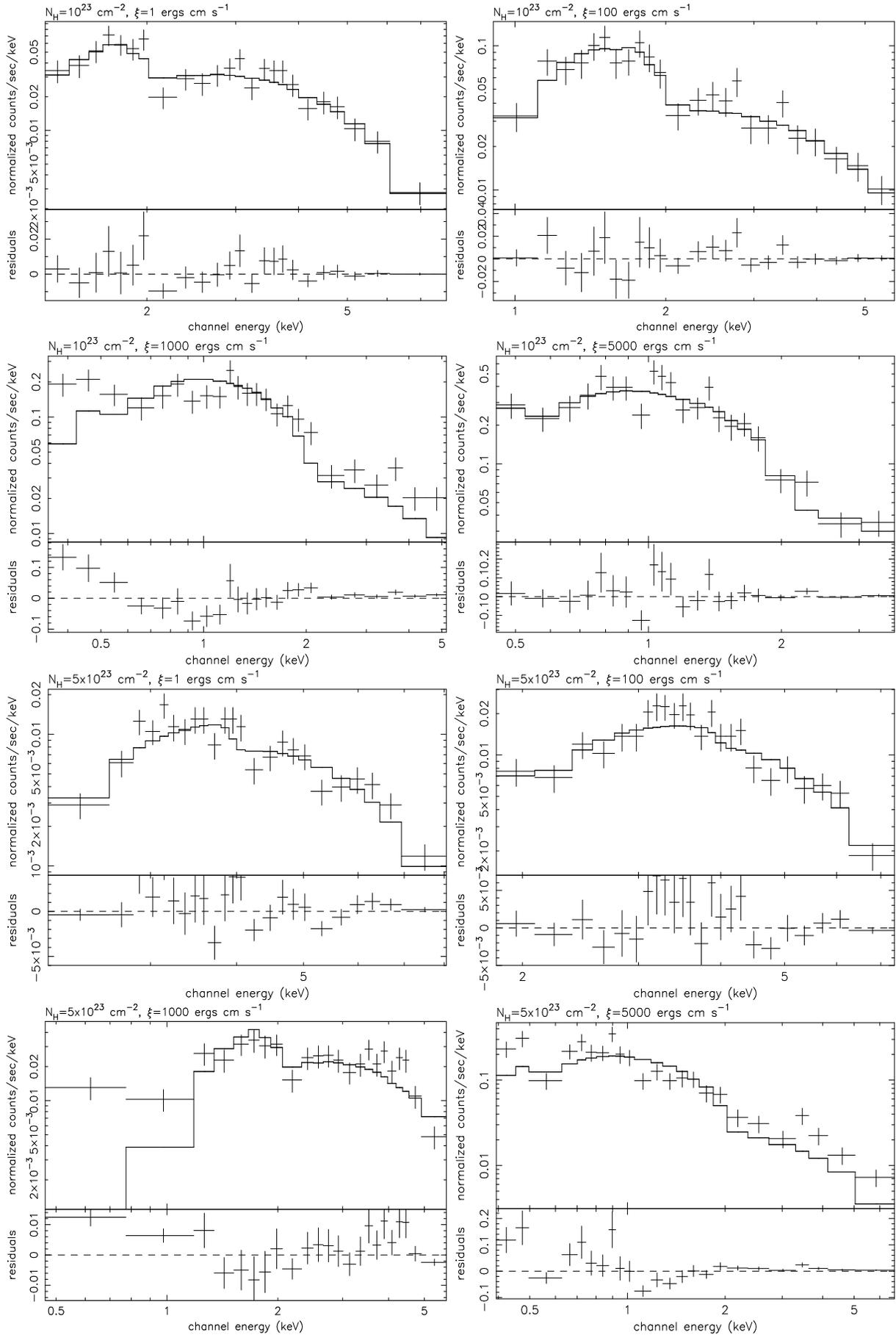

  \centering
  \rotatebox{-90}{\includegraphics[width=6cm, height=8cm]{7791fig1.ps}}
  \rotatebox{-90}{\includegraphics[width=6cm, height=8cm]{7791fig2.ps}}
  \rotatebox{-90}{\includegraphics[width=6cm, height=8cm]{7791fig3.ps}}
  \rotatebox{-90}{\includegraphics[width=6cm, height=8cm]{7791fig4.ps}}
  \rotatebox{-90}{\includegraphics[width=6cm, height=8cm]{7791fig5.ps}}
  \rotatebox{-90}{\includegraphics[width=6cm, height=8cm]{7791fig6.ps}}
  \rotatebox{-90}{\includegraphics[width=6cm, height=8cm]{7791fig7.ps}}
  \rotatebox{-90}{\includegraphics[width=6cm, height=8cm]{7791fig8.ps}}
     \caption{Characteristic  spectra from the high-count simulated sample. 
       We present two different  $N_H$ values, namely 10$^{23}$ and 
       5$\times$10$^{23}$ cm$^{-2}$ in four different ionisation states 
       (i.e. $\xi$=1, 100, 1000 \& 5000).}
    \label{spectra}
    \end{figure*}

Both distributions presented in this figure peak around  $\chi^{2}_{\nu}$$\sim 1$
while the high count distribution is shifted to higher  $\chi^{2}_{\nu}$ values 
in respect to the low count distribution. Furthermore these histograms 
suggest that the $\sim$85 percent of the low count and the $\sim$75 per 
cent of the high count sources would give an acceptable statistical result 
when fitted using a simple absorbed power law model 
(a value of 1 of the  $\chi^{2}_{\nu}$ implies that the adopted model has an 
acceptance probability of 0.4 for 20 degrees of freedom).    
Only in a few cases the cold absorption power-law model could be automatically 
rejected based on the  $\chi^{2}_{\nu}$ value (e.g. $\chi^{2}_{\nu}\sim2$ implies 
that the cold absorption power-law model is rejected at $\sim3\sigma$ confidence level for 
20 degrees of freedom). 
In Fig. \ref{spectra} we present some characteristic high-count simulated spectra. 
Table 1 summarises the spectral fitting results.   

\begin{table}
\centering
\caption{Spectral fitting results for the eight  example simulated spectra 
presenting in Fig \ref{spectra}}
\label{table1}
\begin{tabular}{ccccc}

${N_{H_{in}}}^a$ & $\xi^{b}$ & $\Gamma^c$  & $ {N_{H_{out}}}^d$  &  $\chi^2/d.o.f^e$ \\
$\times10^{22}$ cm$^{-2}$ & ergs cm s$^{-1}$  &  &  $\times 10^{22}$ cm$^{-2}$  &   \\
\hline
10   & 1     &  1.9 &  $ 9.2^{+0.8}_{-0.9}  $  & 18.9/22 \\
10   & 100   &  1.9 &  $ 5.1^{+0.7}_{-0.6}  $  & 18.1/22 \\
10   & 1000  &  1.9 &  $ 0.2^{+0.3}_{-0.2} $   & 37.8/22 \\
10   & 5000  &  1.9 &  $  -  $   	       & 20.9/20 \\
50   & 1     &  1.9 &  $ 47.6^{+15.2}_{-11.8}$ & 25.9/21 \\
50   & 100   &  1.9 &  $ 26.3^{+4.4}_{-4.1}  $ & 23.1/22 \\
50   & 1000  &  1.9 &  $ 9.1^{+1.7}_{-1.5} $   & 57.5/21 \\
50   & 5000  &  1.9 &  $  -  $   	       & 60.1/24 \\
\end{tabular}
\begin{list}{}{}

\item$^a$ Input $N_H$ value for the simulation model \\
\item$^b$ Input $\xi$ value for the simulation model \\
\item$^c$ Input $\Gamma$ value for the simulation model  \\
\item$^d$ Measured $N_H$ assuming a simple power law model  \\
\item$^e$ The best fit reduced $\chi^2$ values \\
\end{list}

\end{table}

The results presented above suggest that mildly ionised ($\xi\sim100$) 
column densities are well described by a simple power law model. 
This is that, given the source count statistic, it is not easy to 
discriminate between a neutral or an ionised model. On the other hand, 
when the  column density is strongly ionised ($\xi\sim1000$) the 
resulting $\chi^{2}_{\nu}$ values are high and thus the power-law models 
not accepted. Note however that in the simulated spectra there is no sign 
of an Fe K$_{\alpha}$ edge consistent with mildly or highly ionised material but only 
a ``soft excess'' like feature at lower energies that could be modelled using a second 
steeper power law model to improve the 
statistical result. In the case of even stronger ionised 
spectra ($\xi=5000$), the spectral fit with a cold absorption power-law model 
is not accepted for those sources with the higher column density
($N_H>5\times10^{23}$ \cunits (see table 1). 

 However, note that these highly ionised (and thus highly luminous) sources 
exist preferentially at higher redshifts and thus 
the low energy absorption features would be shifted outside the 
observed band. For example the mean redshift of the sources
with L$_X$$>$10$^{44}$ \lunits in the combined sample ({\xmm} and {\chandra}) 
of Akylas et al (2006)  is about 1.6. This suggests that 
the ``soft excess'' like features seen in Fig \ref{spectra} 
would be probably shifted outside the observed energy band. 

\subsection{Results}
\begin{figure}
   \centering
   \includegraphics[width=8.0cm]{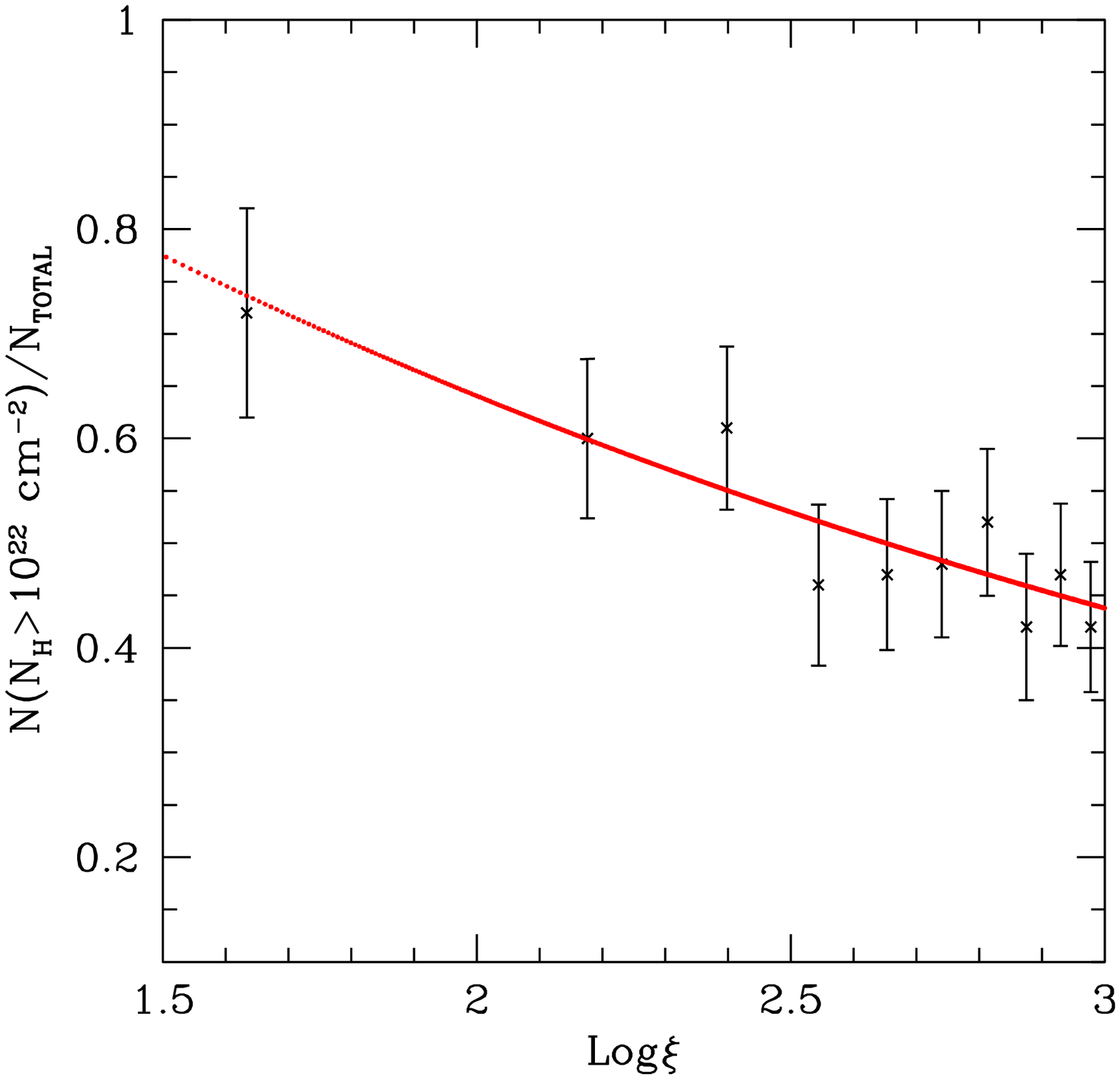}
   \includegraphics[width=8.0cm]{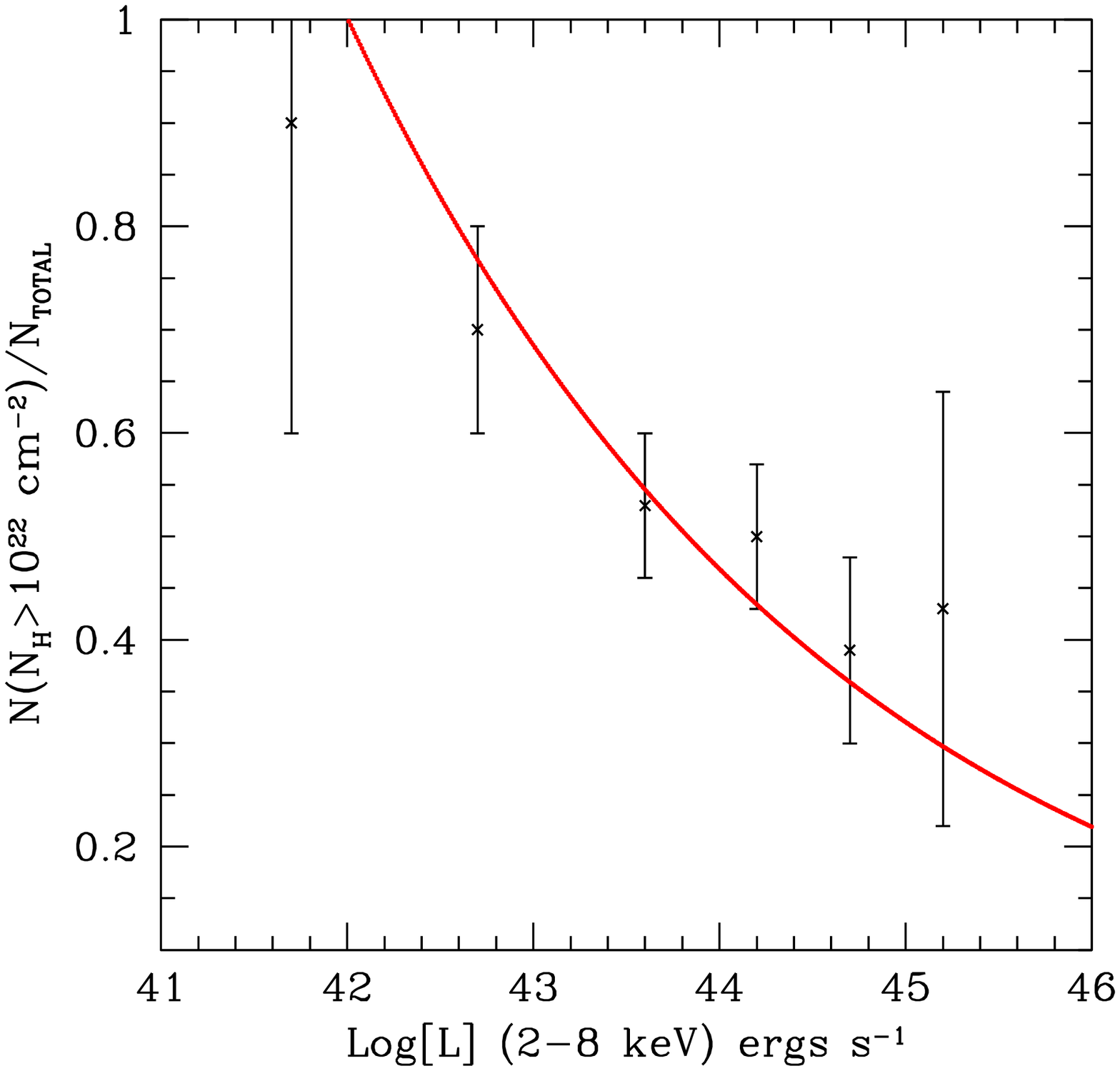}
     \caption{Upper panel: The simulation results for the fraction of 
       obscured AGN as a function of the ionisation parameter (points) 
       together with the best fit power-law model, F$\sim\xi^{-0.16}$ 
       (solid line). Lower panel: The fraction of the obscured AGN as a 
       function of luminosity in the combined {\xmm} and {\chandra} data 
       presented in Akylas et al. (2006) (points). The solid line represents 
       the correlation predicted by our simulation model (F$\sim\L^{-0.16}$).}
    \label{fraction}
    \end{figure}

In the upper panel of Fig. \ref{fraction} we plot the fraction 
of obscured sources in our simulated sample as a function of the 
ionisation parameter. We fit the data with a model of the form 
of F$\sim$$\xi^a$. We find that this fraction drops with increasing 
ionisation parameter as $\xi^{-0.16\pm0.03}$. 
According to the definition  of the ionisation parameter $\xi$, 
the same fraction may be written in terms of luminosity 
as $(L/n_er^2)^{-0.16}$. A similar correlation has been 
suggested by recent studies of {\chandra} and {\xmm}  observations 
(e.g. Akylas et al. 2006; La Franca et al.  2005; Ueda et al. 2003). 
 We fit the data of Akylas et al. (2006) with our 
simulation model $(L/n_er^2)^{-0.16\pm0.03}$ leaving 
 only the normalisation (i.e. $(n_er^2)^{+0.16}$) as a free parameter 
The fit is accepted at the 80 per cent confidence level yielding  
$n_er^2$$\sim$2$\times10^{41}$ cm$^{-2}$ cm$^{-1}$. 
Individual studies of the local bright AGN NGC1068 (e.g. Jaffe et al. 2004) 
and Circinus (e.g. Prieto et al. 2004) show that the obscuring material 
lies outside the BLR up to a distance of a few light years. 
Our estimate for the $n_er^2$ predicts that at these distances 
(1 pc) the density of the material should be $\sim$10$^5$ cm$^{-3}$. 

In Fig. \ref{fraction} (lower panel) we plot the 
{\it observed} fraction of obscured AGN as a function of luminosity
adapted from Akylas et al. (2006), together with the predictions of  
our ionisation  model ($L^{-0.16}$).
Our model predicts that the fraction of obscured AGN reaches 
 the assumed maximum of 80 per cent at a luminosity of 
$4\times10^{42}$ \lunits. 
It is worth noting that the exact shape of the input $N_H$ distribution may 
affect the shape of the Obscured Fraction - $\xi$ relation. This is true 
only for the sources with medium $N_H$ values ie. between 10$^{22}$ 
and 10$^{23}$ cm$^{-2}$. Sources with $N_H<10^{22}$ 
cm$^{-2}$ would always remain lower and sources with $N_H>10^{23}$ 
cm$^{-2}$ would always be higher than the critical value of 10$^{22}$ 
cm$^{-2}$ regardless of the photo-ionisation state of the material.
Our estimations show that if we reduce the number of sources 
with $10^{22}<N_H<10^{23}$ by 50 per cent (see Fig. \ref{nh_sim}) 
we get a flatter Obscured Fraction - $\xi$ relation (F$\sim$$\xi^{-0.1}$).

\section{Discussion}

 \subsection{The warm absorber}
 The results presented above demonstrate that the simplest 
 photo-ionisation model can roughly reproduce   
 the decrease in the fraction of obscured AGN at high 
 luminosities.  If this model is true it would imply that 
 many luminous QSOs would present evidence for  warm absorption. 
 The absorption edges due to OVII and OVIII at 
 0.74  and 0.87 keV respectively   are the strongest signatures 
 of ionised gas at X-ray wavelengths. The high resolution achievable 
 with the grating spectrometers on-board {\xmm} and {\chandra} revealed
 many more absorption features as e.g. the Unresolved  Transitional 
 Array between 0.73 and 0.77 keV (Behar, Sako \& Kahn 2001). 
 These features are blue-shifted and show distinct velocity 
 components (eg. Kaastra et al. 2000). This strongly suggests that 
 the warm absorbers are associated with outflows. {\xmm} high sensitivity 
 spectroscopy with RGS (Guainazzi \& Bianchi 2006) places tight constrains 
 on the location of the warm absorber. These appear to be associated with the 
 molecular torus.  Warm absorbers are found in about 50\% of nearby 
 Seyfert galaxies (Reynolds 1997; Crenshaw, Kraemer \& George 2003). 
 
 The picture was unclear at high (QSO) luminosities.
 are red-shifted at low energies.
 In the {\it ASCA} QSO samples of Reeves \& Turner (2000) and George et al. (2000) 
 there is little evidence for ionised absorption. Because of its high 
 effective area, {\xmm} routinely produced high signal-to-noise spectra and 
 changed drastically the above picture. Piconcelli et al. (2005) find 
 that about 50\% of the 40 QSOs in their PG sample present warm absorbers. 

 It is interesting to investigate whether there is any 
 evidence for the presence of warm absorber features 
 in the spectra of luminous AGN in deep surveys. 
 Here we examine the luminous AGN in the combined sample 
 of Akylas et al. (2006).
 There are about 30 QSOs with $L_X>10^{44}$ {\lunits}  
 below a redshift of $\sim$1.2 where the oxygen absorption edges 
 should be readily detected. Of these only 10 have a good 
 signal-to-noise ratio with over 200 photons in their spectra.
 We fit a warm absorber model and find that 3 sources 
 have an equally good fit  with an ionised $\xi>100$ warm absorber, 
 as compared to the cold absorber. Obviously the limited photon 
 statistics do not allow for a conclusive test in this case. 

 \subsection{The X-ray Baldwin effect}
 The Fe lines above 6 keV provide a census of the state of
 circumnuclear material in AGN. It is therefore reasonable to 
 expect a relation between the obscuration fraction 
 and the strength of the Fe lines. Iwasawa \& Taniguchi 
 (1993) first find that the equivalent width (EW) of the narrow 
 Fe K$\alpha$ at 6.4 keV line decreases with the 2-10 keV X-ray luminosity in 
 a sample of Seyfert galaxies observed by {\ginga} 
 (the so called Iwasawa-Taniguchi or X-ray Baldwin effect).  
  The correlations above closely resemble the well known Baldwin effect
 where the EW of the CIV 1550 {\AA} line decreases with  increasing 
 optical luminosity (Baldwin 1977).  The optical Baldwin effect has been 
 corroborated by the results of Kinney et al. (1990) and more recently 
 Croom et al. (2002) using 2dF data.
 The last authors  find a relation with a slope of -0.128 $\pm$ 0.015.  
 The Baldwin effect can be explained on the basis of either a diminishing 
 covering factor  or alternatively an increasing  ionisation parameter
 (Mushotzky \& Ferland 1984) with increasing luminosity.

 Page et al. (2004) again  find evidence for an X-ray Baldwin effect 
 using Seyfert and QSO data that span more than 5 orders of magnitude 
 in luminosity. They find a correlation between the narrow core of the 
 Fe K$_{\alpha}$ EW and the 2-10 keV X-ray luminosity  
 in the form of $EW\propto L_X^{-0.17\pm 0.08}$. 
 More specifically,  the EW obtains values around 150 eV at low 
 luminosities ($\rm L_{X}(2-10keV)< 10^{43}$ \lunits), dropping to 
 less than 50 eV at high luminosities ($L_X>10^{46}$ \lunits).   
 Jiang et al. (2006) confirmed the X-ray Baldwin effect using {\it Chandra} data.
 Recent results by Bianchi et al. (2007) shed more light to the above issues. 
 They find a highly significant anti-correlation  of narrow Fe K$_{\alpha}$ 
 line EW with respect to the Eddington ratio while no 
 dependence on the black hole mass is apparent. Moreover they find a 
 correlation between the ratio of highly ionised to neutral FeK line flux 
 with luminosity that supports our photo-ionisation scenario. Note however 
 that according to their claims this is a rather weak result 
($\rm Log(R)=(0.12\pm 0.06) + (0.09\pm0.06)Log(L_{X_{44}}$) to explain the 
 observed X-ray Baldwin effect. 

 It is interesting to note that the X-ray Baldwin effect 
 is also present in the broad component of the FeK line.
 Nandra et al. (1997) find an anti-correlation of the broad
 Fe line EW with the 2-10 keV luminosity in a sample 
 of 18 Seyfert galaxies observed by {\it ASCA}. 
 The broad FeK line most likely originates in the accretion disk.
 Nandra et al. (1997) assert that the most obvious explanation 
 for the X-ray Baldwin effect 
 observed is the photo-ionisation of the accretion disk.  


\section{Conclusions} 
 We employ X-ray spectral simulations to determine the effect 
 of photo-ionisation on the obscuring torus in AGN. 
 In particular,  our goal is to explore whether 
 photo-ionisation can explain the steep decrease 
 in the fraction of obscured AGN with increasing 
 luminosity as witnessed in X-ray surveys.    
 Our conclusions can be summarised as follows: 

 \begin{itemize} 
 \item{Photo-ionisation can reproduce the observed decrease  
 in the absorbed AGN fraction as a function of luminosity.
  We find that the fraction of absorbed AGN should fall with luminosity as 
   $L^{-0.16\pm0.03}$ in rough agreement with the observations.} 

  \item{The observed decrease in the obscuring material 
 covering factor is roughly consistent  with the dependence of the 
 Fe K$_{\alpha}$ line equivalent width on luminosity, the X-ray Baldwin effect.}
 
\item{A simple  prediction of the photo-ionised model is that a high fraction 
of  luminous QSOs should present warm absorption features  in their X-ray spectra.
 Future, high effective area mission, such as {\it XEUS} will be able to 
 quantitatively assess this scenario.}
 
 \end{itemize} 
 
\section*{Acknowledgements}
We are grateful to the anonymous referee for a several helpful 
comments and suggestions on this manuscript.

\end{document}